\documentclass{article}

\usepackage[english]{babel}
\usepackage{amssymb,epsf}
\usepackage{epsfig,graphics,indentfirst}

\def\beq{\begin{equation}}
\def\eeq{\end{equation}}
\def\beqa{\begin{eqnarray}}
\def\eeqa{\end{eqnarray}}
\def\bet{\begin{tabular}}
\def\eet{\end{tabular}}
\def\bef{\begin{figure}}
\def\eef{\end{figure}}
\def\half{\frac{1}{2}}

\let\no=\nonumber
\let\noin=\noindent
\let\in=\indent

\def\pr{{\it Phys. Rev.}\ }
\def\prl{{\it Phys. Rev. Lett.}\ }

\def\cqg{{\it Class. Quantum Grav.}\ }

\def\apj{{\it Ap. J.}\ }
\def\aa{{\it Astron. Astrophys.}\ }

\def\aj{{\it Astron. J.}\ }

\begin{document}

\title{Increase of the Number of Detectable Gravitational Waves Signals due to Gravitational Lensing}

\author{M.~Arnaud~Varvella \footnote{Dipartimento di Fisica E.R.~Caianiello, Universit\'a di Salerno - 84081 Baronissi (Sa), Italy and LERMA/ERGA, Universit\'e Paris VI, 4 pl.Jussieu - 75005 Paris, France. e-mail:~varvella@sa.infn.it or varvella@ccr.jussieu.fr}, M.C.~Angonin \footnote{LERMA/ERGA, Universit\'e Paris VI, 4 pl. Jussieu - 75005 Paris, France. e-mail:~m-c.willaime@obspm.fr}, Ph.~Tourrenc\footnote{LERMA/ERGA, Universit\'e Paris VI, 4 pl. Jussieu - 75005 Paris, France. e-mail:pht@ccr.jussieu.fr }}

\date{\today}

\maketitle

\begin{abstract}
This article deals with the gravitational lensing (GL) of gravitational waves (GW). We compute the increase in the number of detected GW events due to GL. First, we check that geometrical optics is valid for the GW frequency range on which Earth-based detectors are sensitive, and that this is also partially true for what concerns the future space-based interferometer LISA. To infer this result, both the diffraction parameter and a cut-off frequency are computed. Then, the variation in the number of GW signals is estimated in the general case, and applied to some lens models: point mass lens and  singular isothermal sphere (SIS profile). An estimation of the magnification factor has also been done for the softened isothermal sphere and for the King profile. The results appear to be strongly model-dependent, but in all cases the increase in the number of detected GW signals is negligible. The use of time delays among images is also investigated.
\end{abstract}

\vspace{0.5cm}

Keyword(s): gravitational waves; gravitational lensing; interferometric detectors of gravitational waves.

\newpage

\baselineskip = 2\baselineskip

\section{Introduction}

Gravitational waves (GW) have been already predicted by A. Einstein
\cite{AE} in 1918, but they have not yet been observed directly because of the weakness of the signal. However, their existence has been indirectly established by the long-term study of the binary pulsar 1913+16 \cite{HT}. The giant interferometers currently under development \cite{geo,ligo,tama,virgo} appear presently to be the most promising GW detectors. They should reach better sensitivities (and over larger bandwidths) than the network of existing resonant bars \cite{igec}, which have already been taking data for years.\\
\in Yet, detecting GW signals will not be straightforward, at least with the first generation of interferometers \cite{CT,LG}. So, any amplification mechanism such as Gravitational Lensing (GL) should be studied accurately to estimate the improvements it could provide.\\
\in GL of electromagnetic radiation has been already studied in details (see e.g. Ref.~\cite{SEF}) and the same formalism can also be applied to GW, because gravitational radiations propagating on a gravitational background are affected in the same way than electromagnetic radiations \cite{T83}.\\
\in This topic has been addressed in the literature by many authors
with various points of view: cosmological waveguides for GW \cite{BCMM}, GW detection using gravitational lenses as detectors \cite{A} and finally, microlensing \cite{DIN} and macrolensing \cite{WST,WB} of gravitational radiation in the high frequency approximation as well as in the diffraction case \cite{R,TN}. In this article we use the same approach as Ref.~\cite{Moriond}.\\
\in GL produces magnified images of GW which could be detected more easily if their magnifications are high enough, it could be detected more easily. Similarly, the magnification effect allows one to explore a larger volume of the Universe, and thus it increases the number of potentially detectable sources. Yet, we will see in the following that the increase in the number of events is limited for the lens models we study. But it depends significantly on the particular model of deflector considered. So, forthcoming papers should study more realistic and sophisticated models to get a more accurate conclusion.\\
\in On the other hand, the successful amplification of one single signal may strongly help a first detection within a not too far future; therefore, studying this problem is important, even if the probability of such a lensing event is small. In addition, it is important to see whether other lensing effects associated to the GW signal amplification (e.g. delays between images) can also be used.\\
\in In the sequel, we analyze the GL effect of GW in the frequency domains which Earth and space-based GW detectors are sensitive to. First, a comparison between Electromagnetic Waves (EMW) and GW is reported. Going to GL of GW, the diffraction limit is then estimated: from this computation, it turns out that geometrical optics can be used for Earth-based detectors and even for LISA, the space-based interferometer project \cite{lisa}, provided that the mass of the deflector is big enough. Then, we compute the increase in the number of GW signals due to GL, and we apply this computation to some lens models. The possibility to use the time delay between two images produced by GL is also investigated. Finally, some conclusions and prospects for future analysis are reported.

\section{Electromagnetic and Gravitational Radiation}

GW \cite{T83,T87} are ripples in the curvature of spacetime, which propagate at the speed of light like EMW. GW are characterized by their wavelength $\lambda _{g}$, much smaller than the radius of curvature of the background space-time. As shown in Tab.~\ref{tab:t1}, EMW and GW are very different on many aspects. In particular, the latter are almost insensitive to matter, which makes them important probes for astronomy \cite{CT,LG}. Moreover, GW detectors are sensitive to the amplitude of the signal -- scaling like 1 / distance -- while EMW are mostly detected through their power, scaling like the square of the distance. Finally, the two frequency ranges are also very different: below few tens of kHz for GW, above tens of million Hz for EMW.\\
\in Yet, as both are waves, we assume that all the formula for the GL of EMW can be used for GW, provided that the geometrical optics approximation is valid. Therefore, we estimate the validity range of this critical assumption in the following section, by computing the diffraction limit parameter and the diffraction cut-off frequency.

\section{Diffraction limit}
\label{sec:limit}

The characteristic angular scale in the GL by a point mass $M_{L}$ is the Einstein angle \cite{SEF}, $\theta _{E}=\left(4GM_{L}D^{-1}/c^{2}\right)^{1/2}$ where $D=D_{OL}D_{OS}/D_{LS}$ is the distance parameter. The various parameters, such as $D_{OL}$, $D_{OS}$, etc., are shown in Fig.~\ref{fig:f1} and defined in the corresponding caption.\\
\in Lensing effects are expected to be significant only when the source, the lens and the observer are aligned within approximately the angle $\theta_E$. When the angular size of the source is greater than $\theta_E$, the relative influence of the lensing is reduced \cite{DW} by dilution. Wave effects in GL of EMW by a point mass $M_L$ depend on the parameter $y$

\beq
 y \; = \; \frac{4 \; \pi \; G}{c^2} \frac{M_L}{\lambda}
   \; = \; 2 \times 10^{4} \left(\frac{ 1 \; m }{\lambda}\right) \left(\frac{M_L}{M_{\odot}}\right)
   \; = \; 6 \times 10^{-5} \left( \frac{\nu}{1 \; Hz} \right) \left(\frac{M_L}{M_{\odot}}\right)
\label{eq:e1}
\eeq

\noindent where $ \lambda $ is the wavelength and $ \nu $ the frequency of the radiation. Using the parameters of a point mass lens, $ y $ can be written as

\beq
 y \; = \; \frac{\pi \; D}{\lambda} \; \theta _{E}^{2} \; = \; \frac{\pi \; \nu \; D}{c} \; \theta _{E}^{2}
\label{eq:e2}
\eeq

\noin where all these quantities have been already defined.\\
\in In terms of the Schwarzschild radius $R_S = 2 G M_L / c^2$:

\beq
 y \; = \; 2 \; \pi \; \frac{R_S}{\lambda} \; = \; 2 \; \pi \; R_S \; \frac{\nu}{c} \quad .
\label{eq:e3}
\eeq

\in The parameter $y$ measures the number of Fresnel zones \cite{DW,DW1} contributing to the lensing effect: when $y \sim \infty $ geometrical optics applies, while for $y \sim 1$ severe effects of diffraction occur and more precise solutions of the wave equation are required.\\
\in Like in another similar computation \cite{ZaB}, we estimate the diffraction limit in the case of GW with this formalism. To distinguish from EMW, the wavelength $\lambda$ becomes $\lambda_g$ and the frequency $\nu$, $\nu_g$. In the broad frequency domain of GW \cite{CT}, GW detection efforts focus on four frequency bands shown in Tab.~\ref{tab:t2}: the extremely low frequency (ELF), the very low frequency (VLF), the low frequency (LF) and the high frequency domain (HF). The probes used to search these GW are the following: the polarization of the Cosmic Microwave Background (CMB) radiation for the first range, the pulsar timing for the second one, the LISA experiment \cite{lisa} for the third one and finally the Earth-based detectors (interferometers and resonant mass) for the last one.\\
\in Fig.~\ref{fig:f2} shows the diffraction parameter evolution in the full range of frequencies listed in Tab.~\ref{tab:t2}. The calculation has been made for two different values of the lens mass $M_L$: $10^{6} \ M_{\odot}$ (massive black hole case) and $10^{9} \ M_{\odot}$ (galaxy case) respectively. Tab.~\ref{tab:t3} shows the values of the diffraction parameter $y$ computed with Eq.~\ref{eq:e1}, the corresponding Einstein radius $R_E=D_{OL} \ \theta_E$ depending on the distances, and the Schwarzschild radius $R_S$ for the two lens masses. For cosmic distances (i.e. $d>>10^{8} m$) the relation $R_{S}\ll R_{E}$ holds true. Therefore, the formula for a point mass lens is valid \cite{AS}.\\
\in From Fig.~\ref{fig:f2} we can deduce that $y\gg 1$ in the frequency range sensitive for Earth-based detectors; so, geometrical optics is valid. This can be true also for LISA: in fact, the region where $y>1$ covers partially the LF range for $M_{L}=10^{9}\ M_{\odot }$ (galaxy case), but for $M_{L}=10^{6}\ M_{\odot }$ (black hole case), one is immediately in the diffraction regime.\\
\in We need to estimate a limit on the mass to be sure that geometrical optics is valid also in this domain. Solving $y=1$ for $\nu =10^{-4}~Hz$ and $1~Hz$ gives $M_{L}\sim M_{max}=2\times 10^{8}~M_{\odot }$ and $M_{L}\sim M_{min}=2\times 10^{4}~M_{\odot }$ respectively (see Eq.~\ref{eq:e1}). When $M_{L}>M_{max}$, geometrical optics is always valid in the LF domain and so in the HF range, while for $M_{L}<M_{min}$, diffraction cannot be neglected. The Black Hole BH Sgr A* is intermediate (see Tab.~\ref{tab:t4}) and so the lens formalism considered here does not apply in the whole GW LF region. Of course, in the HF domain, the geometrical optics approximation extends to smaller masses.\\
\in We also consider another method to compute the diffraction limit in the case of gravitational radiation. If we have a Newtonian gravitational lens, i.e. a lens whose effects can be described by the weak field approximation, we can evaluate the diffraction limit for the lens, i.e. a cut-off frequency, $\omega _{c}$, such as geometrical optics is valid at frequencies higher than $\omega _{c}$, while diffraction effect near a caustic are possible at lower frequencies.\\
\in The cut-off frequency is given by \cite{BH}

\beq
 \omega _{c} \; = \; \left(\frac{\pi}{10} \; \frac{G M_{L}}{c^{3}}\right)^{-1}
             \; = \; \left(\frac{\pi}{5} \; \frac{R_S}{c}\right)^{-1} \quad .
\label{eq:e4}
\eeq

\in Fig.~\ref{fig:f3} shows the decrease of the cut-off frequency $\omega _{c}$ with the mass $M_{L}$. For a mass $M_{L}$ bigger than $10^{6}~M_{\odot }$, the cut-off frequency is lower than $1 \ Hz.$ Considering Tab.~\ref{tab:t2}, it appears that geometrical optics is relevant for Earth-based detectors because the corresponding relevant frequencies are higher than the cut-off. Therefore, an amplification can be expected under conditions similar to the electromagnetic case. In the LF domain, the relevant frequencies for LISA \cite{lisa} are lower than the cut-off for $M_{L}=10^{6}~M_{\odot }$; therefore diffraction effects are expected in suitable conditions for this value of the mass. However for $M_{L}>10^{9}~M_{\odot }$, geometrical optics remains valid for GW potentially detectable in LISA.\\
\in In conclusion, in most cases, geometrical optics is valid in particular for Earth-based detectors. For this reason we will consider in the sequel that this is the case.

\section{Gravitational lensing contribution}
\label{sec:number}

We want to estimate how GL increases the number of detectable GW signals coming from a given direction: if the increase is relevant, GL can be considered as an important tool for GW detection. Calling $N_{0}$ the number of detectable signals in the absence of GL, the presence of a gravitational lens increases this number by a quantity $\Delta N$.\\
\in First, we estimate the number of GW signals detectable without GL effect. To do this, we consider arbitrary sources of GW with wave amplitude $h_{0}$ at a fixed distance $r_{0}$ from the source scaled; let $h_{s}=h_{0}~r_{0}/r_{s}$ be the amplitude at the distance $r_{s}$. We also introduce a threshold, $h_{th}$, to model the detection process: if $h_{s}>h_{th}$, the signal is detected whereas one has a false dismissal in the other case. From the GW amplitude scaling law, this threshold can be converted in a distance $r_{th}$. Knowing the detection threshold $h_{th}$ from GW detectors sensitivity \cite{geo,ligo,tama,virgo,lisa}, we can write the detection condition $h_{s} > h_{th}$ as $r_{s} < r_{th}=(h_{0} / h_{th})~r_{0}$.\\
\in Let $n$ be the GW pulse density, that is the number of signals per unit volume and per year. The total number of detectable sources per year up to a distance $r_{s}$ is $N_{0} = 4 \pi r_{s}^{3} n / 3$, assuming a homogeneous distribution of the source, which is valid at large scale.\\
\in Now, we compute the number of GW signals which can be detected in the presence of GL. GW detectors are sensitive to the signal amplitude rather than to its intensity. The GW amplitude $h$ is proportional to the square root of the energy flux \cite{ST}; therefore the lensed amplitude, $h_{l}$, on the detector is $h_{l}=h_{s}\sqrt{A}$, where $A$ is the magnification factor and $h_{s}$ the unlensed GW amplitude.\\
\in The magnification factor depends on the angle $\beta $ and on the
distance $r_{s}$ of the source, $A=A(\beta ,r_{s})$. The function $A(\beta,r_{s})$ is model-dependent. The condition for the source to be detected is

\beq
r_{s} \; < \; \frac{h_{0}}{h_{th}} \; r_{0} \; \sqrt{A(\beta ,r_{s})} \; = \; \rho (\beta) \quad .
\label{eq:e5}
\eeq

\in Thus, the number of detectable signals in the presence of GL is

\beq
 N \; = \; n \; \int_{0}^{\pi } \left( \int_{0}^{\rho (\beta )}2\pi r^{2}dr\right) \sin \beta \ d\beta
   \; = \; \frac{2}{3} \; n \; \pi \int_{0}^{\pi }\rho ^{3}(\beta ) \ \sin \beta \ d\beta \quad .
\label{eq:e6}
\eeq

\in Using Eq.~\ref{eq:e5}, we finally find

\beq
 N = \frac{N_{0}}{2}\int_{0}^{\pi } \left( \frac{\rho(\beta )}{r_{th}} \right) ^{3} \sin \beta \ d \beta \quad .
\label{eq:e7}
\eeq

\in The relative increase in the number of signals is

\beq
 \frac{\Delta N}{N_{0}} = \frac{1}{2}\int_{0}^{\pi }\left(\left( \frac{\rho (\beta )}{r_{th}}\right) ^{3}-1\right)\sin \beta \ d\beta \quad .
 \label{eq:e8}
\eeq

\in Taking into account Eq.~\ref{eq:e5}, Eq.~\ref{eq:e8} becomes

\beq
 \rho = \frac{h_{0}}{h_{th}} \; r_{0} \; \sqrt{A\left( \beta \right) }\Rightarrow \frac{\Delta N}{N_{0}}
      = \half \int_{0}^{\pi }\left( A\left( \beta \right)^{3/2}-1\right) \sin \beta \,d\beta \quad .
\label{eq:e9}
\eeq

\in If two images occur, one has to take into account separately $A_{+}(\beta ,r)$ and $A_{-}(\beta ,r)$. In this case, the total number of signals is

\beq
 N = N_{+}+ N_{-} = 2N_{0}+(\Delta N)_{+}+(\Delta N)_{-}
\label{eq:e10}
\eeq

\noin where $N_{+}$ and $N_{-}$ are the contributions of the positive and negative part of the magnification factor, each of them corresponding to an image. Thus, the total relative increase is

\beq
\frac{(\Delta N)}{N_{0}}=1+\frac{(\Delta N)_{+}+(\Delta N)_{-}}{N_{0}} \quad .
\label{eq:e11}
\eeq

\in The two relative variations can be computed with Eq.~\ref{eq:e8}.

\section{First lens models comparison}
\label{sec:models}

Different lens models \cite{SEF} can be considered to describe the gravitational sources. We start our analysis with the simplest one, the point mass lens (or Schwarzschild lens); we also study the singular isothermal sphere. In both cases, the increase in the number of signals is computed.

\subsection{Point Mass (Schwarzschild lens) Model}
\label{subsec:point}

This model considers a point mass lens $M_L$: it always gives two images. Notations follow Fig.~\ref{fig:f1} and the GW source is also punctual. We consider the case $D_{LS}/D_{OL} \gg 1$, for which the Einstein angle $\theta _{E}$ becomes independent from the source distance:

\beq
 \theta _{E} \; = \; \sqrt{\frac{4 \; G \; M_L}{c^{2}} \frac{D_{LS}}{D_{OL} \; D_{OS}}}
          \; \approx \; \sqrt{\frac{4 \; G \; M_L}{c^{2}} \frac{1}{D_{OL}}} \quad .
\label{eq:e12}
\eeq

\in So, the magnification factor $A(\beta, r)$ depends only on the angle $\beta$. In this model, one can prove \cite{SEF} that the two contributions to the amplification are

\beq
 A_{\pm} \; = \; \frac{u^{2}+2}{2 \; u\sqrt{u^{2}+4}} \; \pm \; \half
\label{eq:e13}
\eeq

\noin with $u = \beta / \theta _{E}$.\\
\in Looking at Eq.~\ref{eq:e8} and  Eq.~\ref{eq:e12}, the relative increases of the number of signals are

\beq
 \frac{(\Delta N)_{\pm}}{N_{0}} \; = \;
  \frac{\theta _{E}}{2} \; \int_{0}^{\pi /\theta_{E}} \left(\left(\frac{u^{2}+2}{2u\sqrt{u^{2}+4}} \pm \frac{1}{2} \right)^{3/2}-1 \right) \sin \ (u \ \theta _{E}) \ du   \quad .
\label{eq:e14}
\eeq

\subsection{SIS (Singular Isothermal Sphere) Profile}
\label{subsec:sis}

This model uses as lens a sphere of radius $ R $ with a mass distribution $M(R)$ confined in this volume \cite{SEF}. The velocity dispersion $\sigma_v$ scales with the rotational velocity $v_{rot}$ as $v_{rot} = \sqrt{2} \sigma_v$. This model gives multiple images only if the source lies inside the Einstein ring, i.e. for $ \beta < \theta_E$; if the source lies outside the Einstein ring, i.e. for $ \beta > \theta_E$, there is only one image. For this model the positive and negative contribution to the magnification factor are

\beq
 A_{+} \; = \; 1+\frac{1}{u} \qquad  A_{-} \; = \; \left| 1-\frac{1}{u}\right|
\label{eq:e15}
\eeq

\noin where $u=\beta /\theta _{E}$.\\
\in We consider again the case $D_{LS} / D_{OL} \gg 1$. For a given lens distance, the Einstein angle $\theta _{E}$ becomes independent of the distance $D_{LS}$ and is indeed constant

\beq
 \theta_E = \frac{4 \pi \sigma^2_v}{c^2} \frac{D_{LS}}{D_{OS}}
\sim\frac{4 \pi \sigma^2_v}{c^2} \quad .
\label{eq:e16}
\eeq

\in Again, the magnification factor $A(\beta, r)$ depends only on $\beta$; from Eq.~\ref{eq:e8}, we obtain

\beq
\frac{(\Delta N)_{\pm }}{N_{0}}=\frac{\theta _{E}}{2}\int_{0}^{\beta _{\pm
}/\theta _{E}}\left( \left| 1\pm \frac{1}{u}\right| ^{3/2}-1\right) \sin
(u\theta _{E})\ du
\label{eq:e17}
\eeq

\noin where $\beta _{+}=\pi $ and $\beta _{-}=\theta _{E}$ because
$(\Delta N)_{-}/N_{0}$ depends only on the sources for which $\beta < \theta_{E}$.

\subsection{Applications to possible lens candidates}

It is interesting to estimate the increase of the number of signals computed in Sec.~\ref{sec:number} for the two particular models we have considered above using three hypothetical sources whose mass and distance are respectively the mass and the distance of the Virgo cluster, of a typical galaxy at $1$ Mpc and of the black hole BH Sgr A* in the center of the Galaxy.\\
\in Tab.~\ref{tab:t4} shows the relative variation of the number of GW signals for these different examples. It is clear that the contribution due to the GL effect is really negligible. However the results are strongly model-dependent. The dependence on the model is emphasized in Fig.~\ref{fig:f4} where it is clear that the results from the SIS profile are better than the results from the Schwarzschild lens by several orders of magnitude. From Fig.~\ref{fig:f4}, one can also notice that the more aligned the source with the observer and the lens the higher the magnification factor.

\section{Other interesting lens profiles}

We now consider two other lens profiles used by several authors to describe the mass distribution of the deflector: the softened isothermal sphere and the generalized King profile.

\subsection{The softened isothermal sphere}

The softened isothermal sphere \cite{NB} is more complex than the SIS profile. The mass distribution includes a characteristic core of radius $r_{c}$.\\
\in Let us define $\theta _{c}=r_{c}/D_{OL}$, the core angular position. Then the magnification factor for this model is

\beq
 A_{\pm} \; = \; 1 \pm \frac{D_{LS}}{D_{OS}} \frac{4 \; \pi \; \sigma^2}{c^2} \frac{1}{\beta} \frac{\theta}{\left( \theta_c^2 + \theta^2 \right)^{1/2}}
\label{eq:e18}
\eeq

\noin where $\theta $ is the image position given in Fig.~\ref{fig:f1}, while $\sigma$ has the same meaning as $\sigma _{v}$ in the SIS profile above. When $\theta _{c}=0$ we obtain the SIS profile.\\
\in Fig.~\ref{fig:f5} shows the estimation of the magnification factor for a lens with the softened isothermal sphere profile: the computation has been made for the three different lens candidates already considered for the Schwarzschild lens and the SIS profile. The magnification factor depends on the ratio $x=\theta_c/\theta$.
One can remark that the higher the lens mass, the stronger the amplification.

\subsection{The generalized King profile}

The second model we consider in this section is a generalization of the King profile \cite{Golse,K}. The mass distribution family is characterized by a core radius $r_c$ and an exponent $\alpha$, for which values $1/2$ and $0$ have been considered.\\
\in The magnification factor is

\beq
 A \; = \; \left(1-2 \frac{\Sigma}{\Sigma_0}+\frac{\Sigma^2}{\Sigma_0^2}-\half \; \frac{\Sigma_0 \Sigma}{\Sigma_c^2} \; \frac{(1+2 \alpha)^2 x^4}{2+(1-2 \alpha)x^2} \right)^{-1}
\label{eq:e19}
\eeq

\noin where $\Sigma$ is the mass density depending on the distance $R$; the quantities $\Sigma_0$ (the core mass density projected on the lens plan), $I_{1+\alpha}$ and $\Sigma_c$ (the critical density), are defined below

\beqa
 \Sigma_0 \; &=& \; \frac{4}{3} \; I_{1+\alpha} \; \rho_0 \; r_c \; , \qquad \Sigma_{c} \; = \; \frac{c^2}{4 \; \pi \; G} \frac{D_{OL} \; D_{LS}}{D_{OS}} \no \\
 I_{1+\alpha} &=& \int_0^{+\infty} \frac{dx}{(1+x^2)^{1+\alpha}} \quad .
\label{eq:e20}
\eeqa

\in Fig.~\ref{fig:f6} shows the estimation of the magnification factor for a lens with the generalized King profile: the computation has been made considering Virgo-like clusters as lens and for two particular values of the exponent $\alpha$: $0$ and $1/2$. As we can see, the best amplification is obtained with $\alpha= 1/2$. The magnification factor depends on the ratio $x=R/r_c$. The relative increases of the number of signals can be computed using Eq.~\ref{eq:e8}. Yet, under reasonable assumptions, the computed increases are very small.

\section{Twin signals}

Two images of the same lensed source reproduce an identical GW signal. This could be very helpful for a first GW detection for which the signal amplitude does not exceed significantly the noise level. One can distinguish two kinds of twin signals. In the first case, an angular separation between the two images can be achieved: the twin signals correspond to two directions in the sky. On the other hand, when no angular separation is achieved, only a time delay exists between the two images: this is the second case.\\
\in Burst sources can be detected by GW detectors in coincidence, for instance between Virgo and the LIGO interferometers. Periodic signals can be detected by a single detector using their periodicity. Rough calculations show that an angular separation could happen for periodic signals, observed over one year in a single detector, but not for GW pulses in a network of Earth-based detectors. In both cases, the gravitational signal will be the repetition of two identical signals (or more if the lens gives more than two images), coming from the same direction.\\
\in Tab.~\ref{tab:t5} displays some orders of magnitude for the delay between the two signals: the source position angle $\beta$ is normalized to $60~''$. Yet, uncertainties on $\beta$ are so large that one cannot estimate accurately the delay without additional information on the source.

\subsection{Detectability}

In principle, the analysis of Sec.~\ref{sec:number} can be carried out for both Earth-based and Space-based detectors. Tab.~\ref{tab:t6} shows the values of the foreseen sensitivities (at the frequency of $1~kHz$) of Earth-based GW interferometers presently working or under construction; for each of them the arm-length is also given as it is a key parameter for the interferometer final sensitivity. In case of LISA, the threshold for the amplitude is different because the GW frequency range is lower and the expected detectable sources are different: $h_{th}$ has a value $10^{-23}$ at $10^{-3} \ Hz$ for an integration time of 1 year and an isotropic average over source directions.\\
\in In order to observe two or more images, the weakest has to be detected, therefore the detectability condition becomes

\beq
 h_{s}\sqrt{A_{-}} \gtrsim h_{th} \quad .
\label{eq:e21}
\eeq

\in If the two signals have to be detected by the four interferometers quoted in Tab.~\ref{tab:t6}, one must choose for common threshold $h_{th}$ the value $h_{m}$ giving the worst sensitivity.\\
\in Once more, considering the two lens models presented in Sec.~\ref{sec:models}, gives disappointing results. For the Schwarzschild lens, in the case $u<0.5$ (i.e. $\beta <\theta_{E}/2$) one finds $\sqrt{A_{-}}>0.8$ and thus $h_{s}\gtrsim 1.25 \ h_{m}$. Even if the images could be detected, this result does not change dramatically the order of magnitude of the amplitude for which GW which can be observed. Moreover, the number of such twin signals will be extremely low because the GW corresponding sources must lie in the solid angle $\left( \theta _{E}/2\right) ^{2}$, where $\theta_{E}\ll 1$.\\
\in The result is very similar for the SIS profile. We have outlined that in this case the condition $\beta < \theta _{E}$ must hold true in order to have two images (see Sec.~\ref{subsec:sis}). This is a necessary condition for the signal to be amplified. Therefore the corresponding sources must lie in the very small solid angle
$\left( \theta _{E}\right)^{2}$. So, the number of such signals will be dramatically low.\\
\in On the other hand, if we want to observe a large part of the sky, we must choose for instance $u\lesssim 1/\theta _{E}$ (i.e. $\beta _{E}\lesssim 1$ rad). One calculates $\sqrt{A_{-}}\lesssim \theta _{E}^{2}$ and therefore $h_{s}\gtrsim h_{m}/\theta _{E}^{2}$. This is a so high amplitude that there is no chance that such a signal exists.

\section{Conclusions}

We analyzed GL effects on gravitational radiation. First, we showed that the diffraction is negligible for Earth-based GW detectors (and for LISA in a limited GW frequency range depending on the lens mass) and that geometrical optics is relevant in our analysis. This is true for a point mass lens, but it has to be verified for lenses with a different mass distribution.\\
\in We showed it is possible to compute precisely the relative increase of the number of GW signals due to GL for a single deflector. We performed the calculation for two simple lenses, the Schwarzschild lens and the singular isothermal sphere: the variations obtained are negligible for both models. This computation proved also that for similar lenses (with same mass and same distance to the observer) the results are clearly model dependent (see Fig~\ref{fig:f4}). Some other characteristic profiles of the mass distribution have also been considered: the softened isothermal sphere and the generalized King profile for which the magnification factor is directly computed. Lens candidates used for numerical computations are Virgo-like clusters (same mass and distance), galaxies at $1~Mpc$ and BH-like Sgr A*.\\
\in Results obtained suggest the necessity to repeat in the future the same analysis with more realistic and sophistical lens models, for instance the elliptical ones \cite{BK,KB}; moreover, better results could be achieved considering a system of deflectors instead of a single one, averaging on the value of each single amplification.\\
\in Time delay could be an useful tool to detect periodical GW sources: yet, a classification of these sources is necessary to make this possibility more realistic.\\
\in According to the results obtained in this article, we can affirm that the GL of GW is not statistically interesting and it will not contribute significantly to the new astronomy based on the observation of GW. However, our results came out from the hypothesis that GW interact in the same way of EMW with the matter; this is only a first approach of the problem. The study of exceptional situations remains useful: in fact, the estimation of the GL effect due to the interaction of GW with matter could produce better GW amplifications as GW can pass through the matter and are very few absorbed. This will be the aim of future investigation.

\baselineskip = 0.5\baselineskip

\newpage

\begin{table}[h!]
\begin{center}
\bet{|c|c|c|} \hline
 & Electromagnetic Radiation & Gravitational Radiation \\ \hline
Nature & EM fields through space-time & Geometry of space-time \\
Source & Incoherent superpositions of particles & Mass-energy coherent motion \\
Wavelength & Small compared to sources & Comparable to \\
 & & or bigger than sources \\
Properties & Easily absorbed, scattered, dispersed & Nearly unperturbed by matter \\
Frequency range & Above $10^7$~Hz & $\left[ 10^{-18} \, \mathrm{Hz} \, ; \,  \mathrm{\leq few~tens~of~kHz} \right]$ \\
Detectable quantity & Power & Amplitude \\
\hline
\eet
\end{center}
\caption{Differences between electromagnetic and gravitational radiation.\label{tab:t1}}
\end{table}

\newpage

\begin{table}[h!]
\begin{center}
\bet{|c|c|c|} \hline
Frequency [Hz]          & Range & Probes             \\ \hline
 $10^{-18} \div 10^{-15}$ & ELF   & CMB \ radiation  \\
 $10^{-9}  \div  10^{-7}$ & VLF   & Pulsar timing    \\
 $10^{-4}  \div  1      $ &  LF   & LISA experiment  \\
 $1        \div  10^{4} $ &  HF   & Earth-based \ detectors
 \\ \hline
\eet
\end{center}
\caption{Explored GW frequency ranges and corresponding probes.\label{tab:t2}}
\end{table}

\newpage

\begin{table}[h!]
\begin{center}
\bet{|l|c|c|c|c|} \hline
          &  $M_L~[M_{\odot}]$&$R_E~[m]$&$R_S~[m]$ & $y$\\ \hline
black hole&$10^6 $&$7 \times 10^{4} \ d$&$3 \times 10^{9} $&$ 60 \ \nu_g$  \\
  galaxy  &$10^9 $&$4 \times 10^{4} \ d$&$3 \times 10^{12}$&$6 \times 10^{5} \ \nu_g $ \\ \hline
\eet
\end{center}
\caption{Characteristic values for two examples of lens, with
$d = \left( D_{OL}  D_{LS} / D_{OS} \right)^{1/2}$.\label{tab:t3}}
\end{table}

\newpage

\begin{table}[h!]
\begin{center}
\bet{|c|c|c|c|c|c|c|} \hline
 Model&Lens &$M_L$&$D_{OL}$&$\theta _{E}$&$\sigma_v$&$\left(\frac{\Delta N}{N_{0}}\right)$  \\
   &&[$M_{\odot}$]&[Mpc]&["]&[km/s]& \\ \hline
 Point &Virgo-like cluster& $10^{14}$&$15$&$231$&&$5 \times 10^{-7}$ \\
 mass&galaxy-like&$10^{9}$&$1$&$2.8$&&$10^{-10}$ \\
 &BH-like Sgr A*&$2.6 \times 10^{6}$&$8 \times 10^{-3}$&$1.6$&&$3 \times 10^{-11}$ \\ \hline
 SIS&Virgo-like cluster &$10^{14}$& $15$ & $14.1$ &$700$&$10^{-4}$ \\
profile&galaxy-like&$10^{9}$& $1$&$ 1.1$ &$200$&$8 \times 10^{-6}$ \\
&BH-like Sgr A*&$2.6 \times 10^{6}$&$8 \times 10^{-3}$ & $0.6$&$150$&$4 \times 10^{-6}$ \\ \hline
\eet
\end{center}
\caption{Comparison between two lens models: the Schwarzschild lens and the SIS profile.\label{tab:t4}}
\end{table}

\newpage

\begin{table}[h]
\begin{center}
\bet{|c|c|c|}
\hline
Model & Lens  & $\Delta t$ \\
&     &        [($\beta/60'') $ \ years] \\ \hline
Point & Virgo-like cluster & $1.2\times 10^{5}$ \\
mass  & galaxy-like & $120$ \\
& BH-like Sgr A* & $0.3$ \\ \hline
SIS & Virgo-like cluster & $6\times 10^{3}$ \\
profile & galaxy-like & $37$ \\
& BH-like Sgr A* & $0.2$ \\ \hline
\eet
\end{center}
\caption{Time delay comparison between two lens models: the Schwarzschild lens
and the SIS profile.\label{tab:t5}}
\end{table}

\newpage

\begin{table}[h!]
\begin{center}
\bet{|c|c|c|}
\hline
Interferometer & Arm-Length & Threshold Amplitude $h_{th}~@~1~kHz$ \\
& [m]  & [${\mathrm{Hz}}^{-1/2}$] \\ \hline
VIRGO  & $3000$& $3 \times 10^{-23}$ \\
LIGO   & $4000$ &$1 \times 10^{-22}$ \\
GEO600 & $600$& $2 \times 10^{-22}$ \\
TAMA300 & $300$ & $5 \times 10^{-21}$ \\ \hline
\eet
\end{center}
\caption{Threshold amplitudes for Earth-based interferometers.}
\label{tab:t6}
\end{table}

\newpage

\bef
  \centering{
 \epsfysize=10cm
 \epsffile{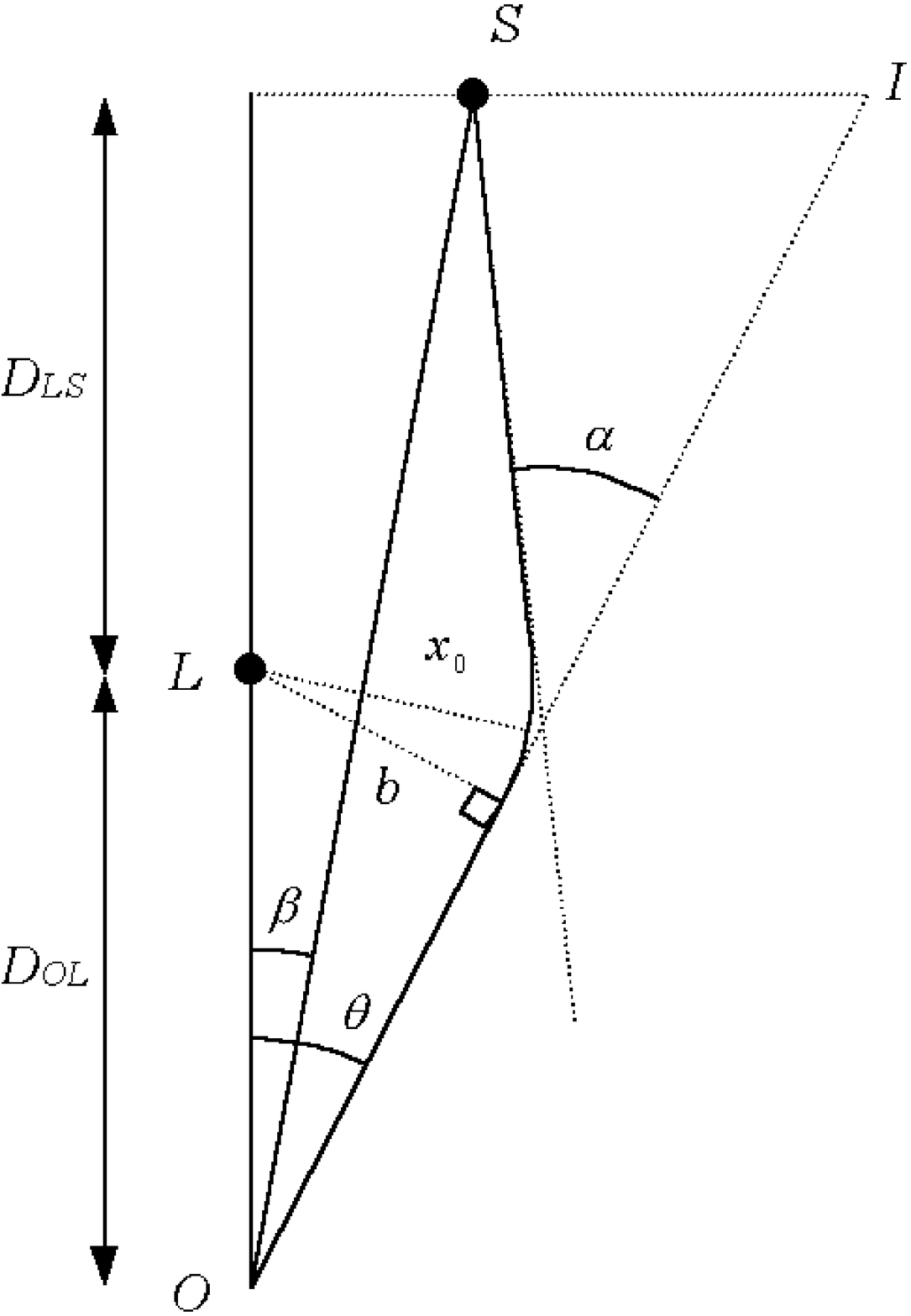}}
 \caption{Lensing diagram for a point mass lens L. S is the source at angle $\beta$ and I its actual image deflected at angle $\alpha$, $D_{LS}$ is the distance between the lens and the source, $D_{OL}$ the distance between the observer and the lens and $D_{OS}$ the distance between the observer and the source.\label{fig:f1}}
\eef

\newpage

\bef
 \centering{
 \epsfysize=10cm
 \epsffile{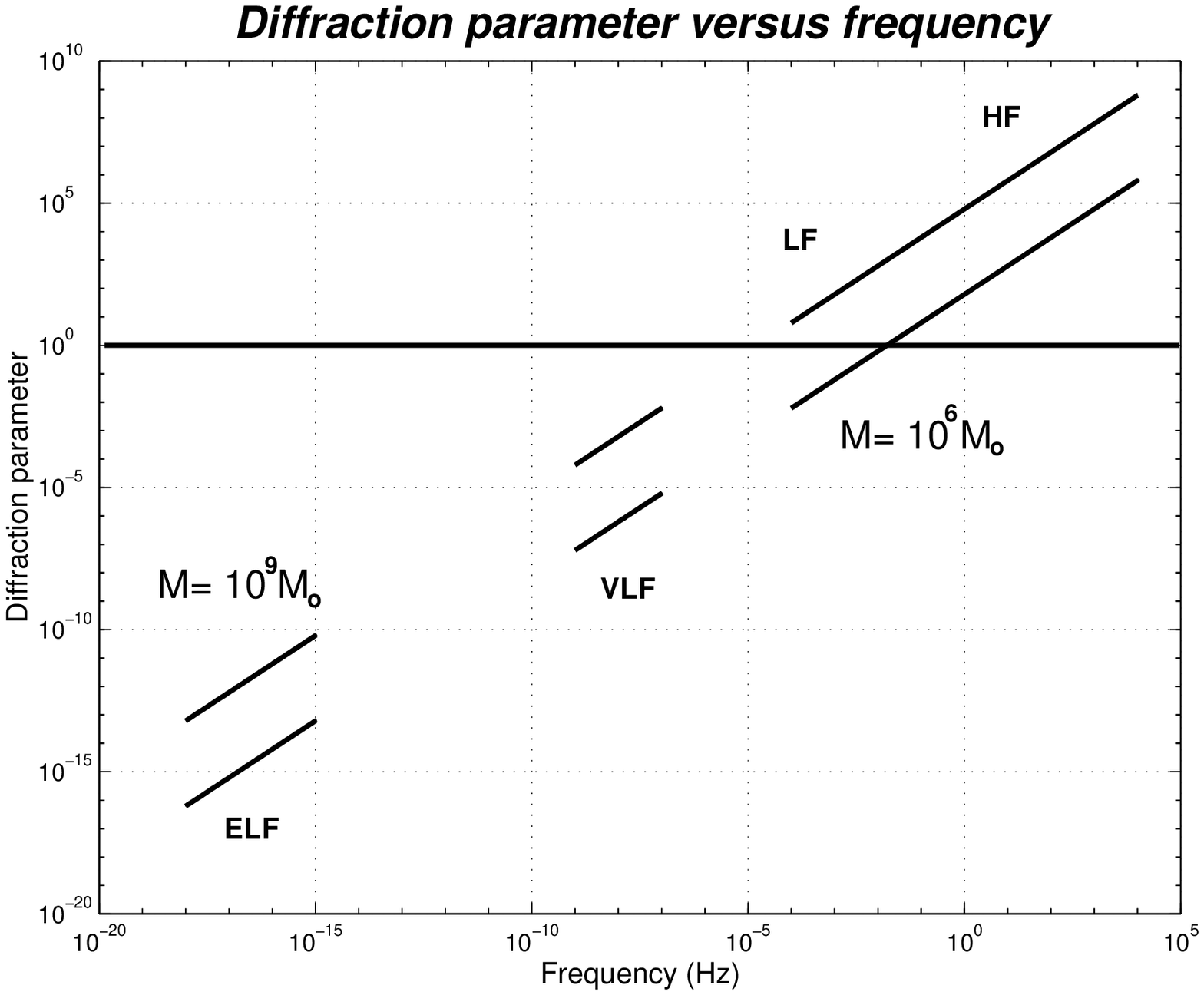}}
 \caption{Diffraction parameter versus GW frequency range for two different values of the lens mass $M_L$: $10^{6} \ M_{\odot}$ (black hole case) and $10^{9} \ M_{\odot}$ (galaxy case) respectively. In the frequency range sensitive for Earth-based detectors $y\gg 1$, so geometrical optics is valid. In the case of LISA, the region where $y>1$ covers partially the LF range for the galaxy case, but for the black hole case, one is immediately in the diffraction regime. Note that both scales are logarithmic.\label{fig:f2}}
\eef

\newpage

\bef
 \centering{
 \epsfysize=10cm
 \epsffile{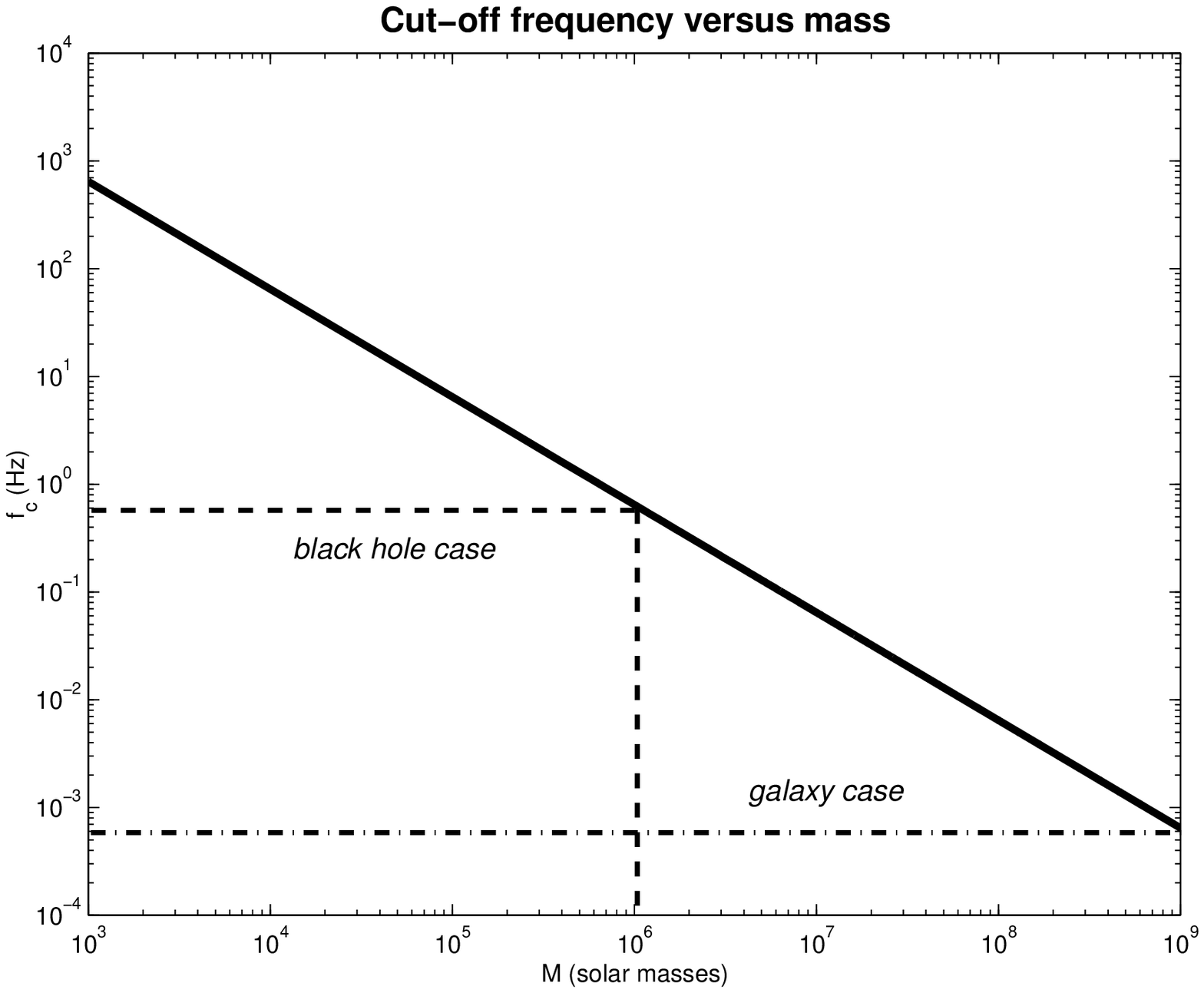}}
 \caption{Cut-off frequency $\omega_c$ versus lens mass $M_{L}$. For a mass $M_{L}>10^{6}~M_{\odot }$ (black hole case), the cut-off frequency is lower than $1~Hz$: geometrical optics is relevant for Earth-based detectors because the corresponding relevant frequencies are higher than the cut-off. In the LF domain, the relevant frequencies for LISA are lower than the cut-off for the black hole case, but for $M_{L}>10^{9}~M_{\odot }$ (galaxy case), geometrical optics remains valid for GW potentially detectable in LISA. Note that both scales are logarithmic.\label{fig:f3}}
\eef

\newpage

\bef
 \centering{
 \epsfysize=10cm
 \epsffile{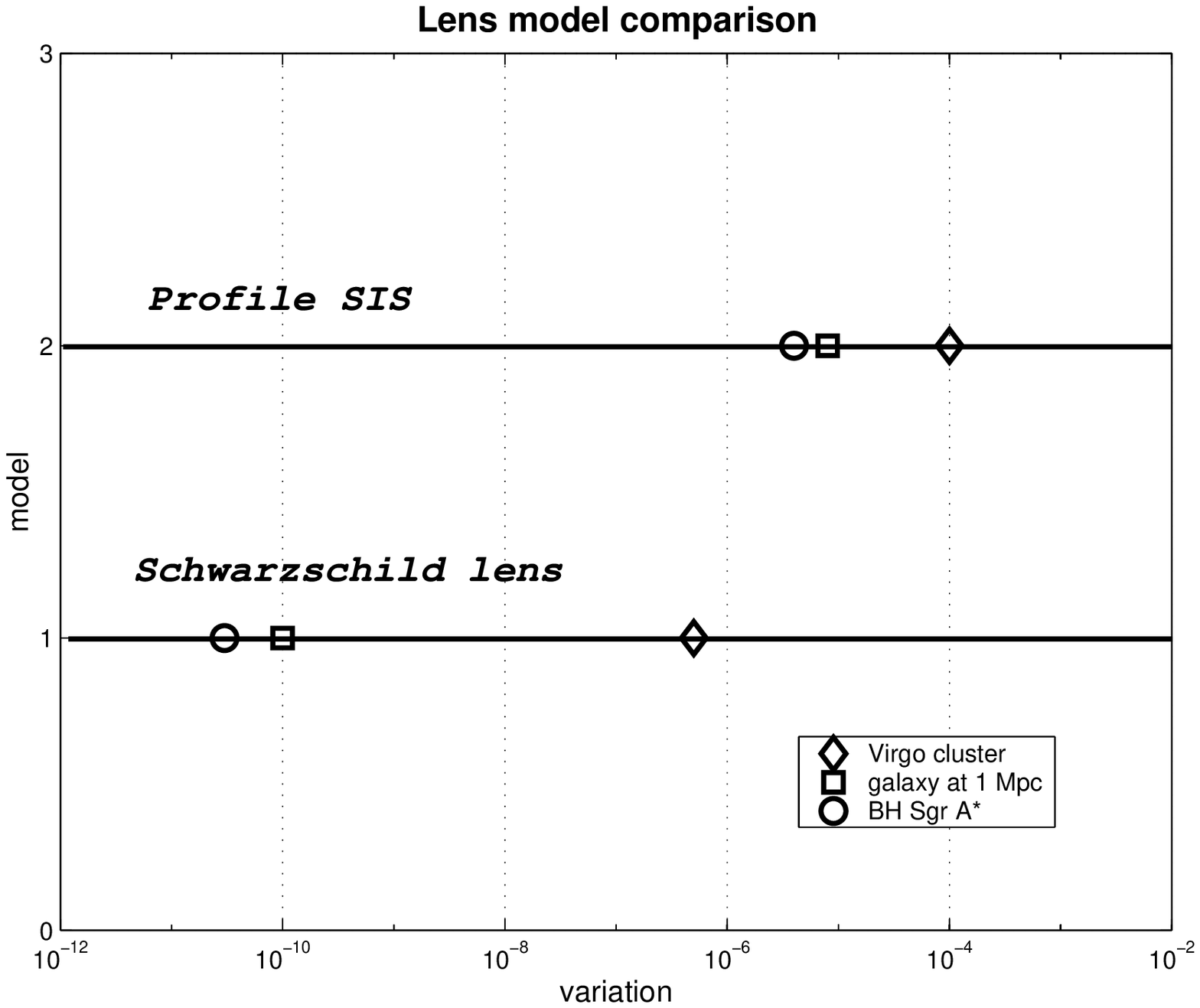}}
 \caption{Model dependence evidence for two different lens models: from the bottom, the first one is the Schwarzschild lens and the second one is the SIS one. Each lens model has been considered with the three lens candidates: the Virgo-like cluster, a galaxy-like at $ 1 \ Mpc $ and the Black Hole-like BH Sgr A*. Results from the SIS profile appear to be better than ones from the Schwarzschild lens by several orders of magnitude.\label{fig:f4}}
\eef

\newpage

\bef
 \centering{
 \epsfysize=10cm
 \epsffile{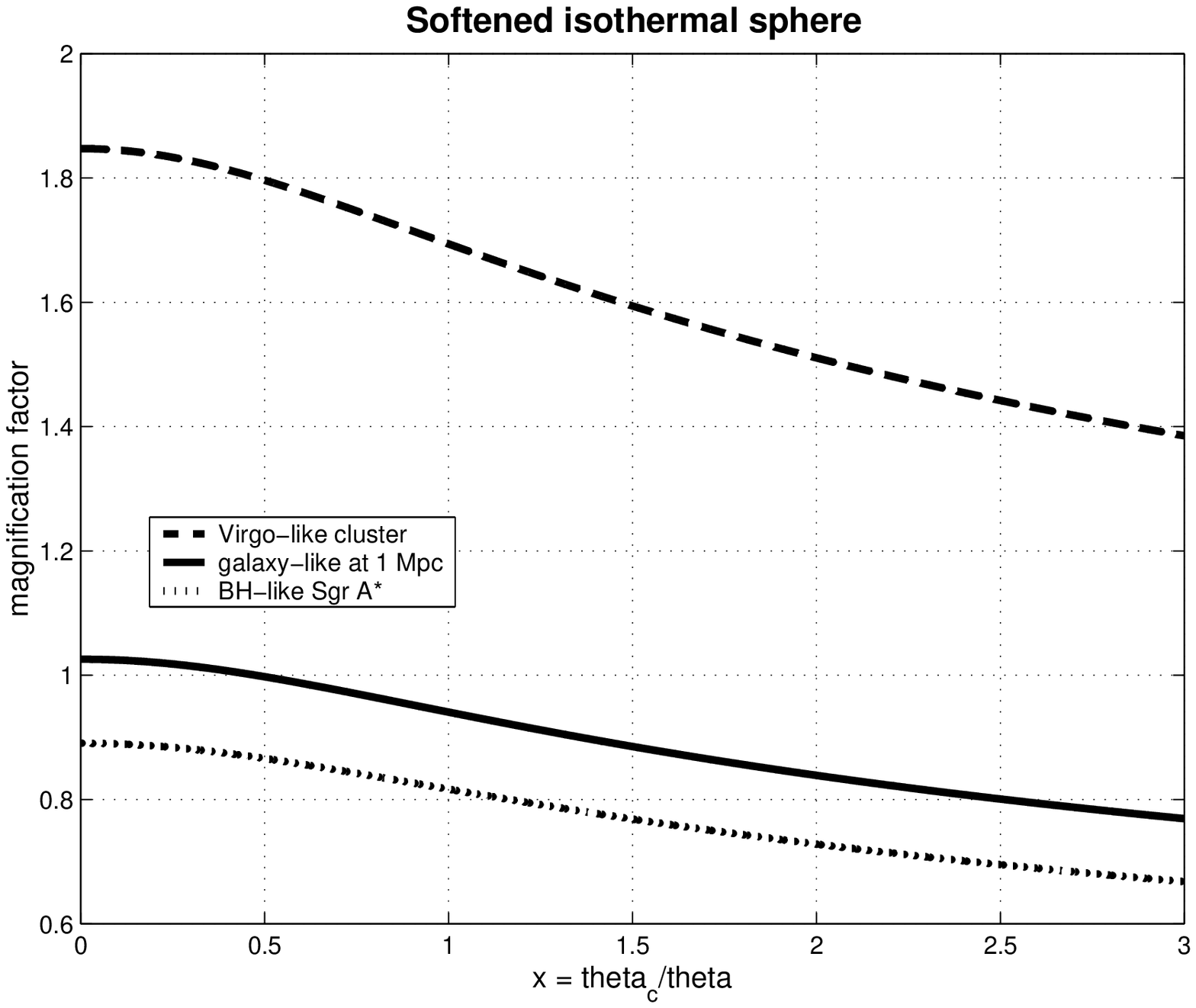}}
 \caption{Magnification factor estimation for the softened isothermal sphere profile in the case of three lens candidates: the Virgo-like cluster, a galaxy-like at $ 1 \ Mpc $ and the BH-like Sgr A*. The magnification factor is showed as a function of the ratio $x$ depending on the core radius of the lens mass.\label{fig:f5}}
\eef

\newpage

\bef
 \centering{
 \epsfysize=10cm
 \epsffile{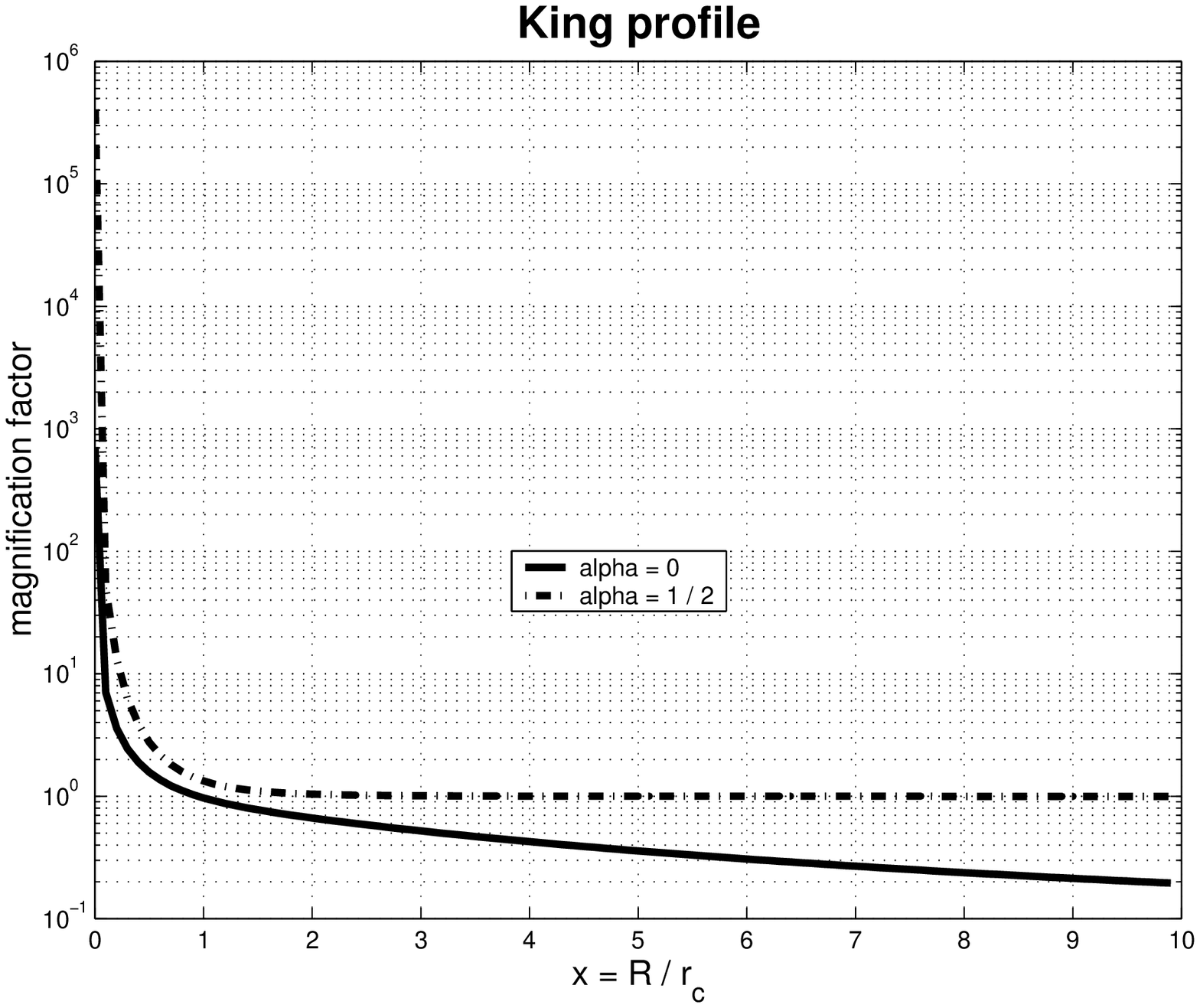}}
 \caption{Magnification factor comparison between two models of the generalized King profile obtained with two particular values of the exponent $\alpha$: $0$ and $1/2$, choosing the Virgo-like cluster as lens. The magnification factor is showed as a function of the ratio $x=R/r_c$ depending on the core radius of the lens mass. The best amplification is obtained with $\alpha= 1/2$.\label{fig:f6}}
\eef

\end{document}